\documentclass[sigconf]{acmart}
\AtBeginDocument{%
  }

\setcopyright{acmlicensed}
\copyrightyear{2018}
\acmYear{2018}
\acmDOI{XXXXXXX.XXXXXXX}
\acmConference[Conference acronym 'XX]{Make sure to enter the correct
  conference title from your rights confirmation email}{June 03--05,
  2018}{Woodstock, NY}
\acmISBN{978-1-4503-XXXX-X/2018/06}

\begin{document}

\title{Large Language Model Few-Shot Prompting with Dilemma Training Outperforms Human Surrogates in Predicting Patient Preferences}


\author{Natasha Ureyang}
\email{natashau@nus.edu.sg}
\affiliation{
  \institution{Telehealth Core, National University of Singapore}
    \city{Singapore}
  \country{Singapore}
}
\author{Sebastian Porsdam Mann}
\email{sebastian.porsdam.mann@jur.ku.dk}
\affiliation{Centre for Advanced Studies in Bioscience Innovation Law, Faculty of Law
  \institution{University of Copenhagen}
    \city{Copenhagen}
  \country{Denmark}
}

\author{Yuxin Liu}
\email{yuxinliu@nus.edu.sg}
\affiliation{
  \institution{Centre for Biomedical Ethics, National University of Singapore}
    \city{Singapore}
  \country{Singapore}
}

\author{Zuriel Hassirim}
\email{zuriel.hassirim@u.nus.edu}
\affiliation{
  \institution{Centre for Biomedical Ethics, National University of Singapore}
    \city{Singapore}
  \country{Singapore}
}

\author{Melanie Almonte}
\email{m.almonte@imperial.ac.uk}
\affiliation{
  \institution{Imperial College London, Epidemiology and Biostatistics, School of Public Health}
    \city{London}
  \country{United Kingdom}
}

\author{Wenhao Chen}
\email{wenhaochen@nus.edu.sg}
\affiliation{
  \institution{Telehealth Core, National University of Singapore}
    \city{Singapore}
  \country{Singapore}
}

\author{Joyce Ng}
\email{e1526387@u.nus.edu}
\affiliation{
  \institution{Telehealth Core, National University of Singapore}
    \city{Singapore}
  \country{Singapore}
}

\author{Thant Nay Lin}
\email{ephv762@nus.edu.sg}
\affiliation{
  \institution{Telehealth Core, National University of Singapore}
    \city{Singapore}
  \country{Singapore}
}

\author{Aung Thiha}
\email{ephv771@nus.edu.sg}
\affiliation{
  \institution{Telehealth Core, National University of Singapore}
    \city{Singapore}
  \country{Singapore}
}

\author{Gerald CH Koh}
\email{ephkohch@nus.edu.sg}
\affiliation{
  \institution{Telehealth Core, National University of Singapore}
    \city{Singapore}
  \country{Singapore}
}

\author{Brian David Earp}
\authornote{Both authors contributed equally as senior authors to this research.}
\email{bdearp@nus.edu.sg}
\affiliation{
  \institution{Centre for Biomedical Ethics, National University of Singapore}
  \city{Singapore}
  \country{Singapore}
  }
\author{Pin Sym Foong}
\authornotemark[1]
\email{pinsym@nus.edu.sg}
\affiliation{
  \institution{Telehealth Core, National University of Singapore}
  \city{Singapore}
  \country{Singapore}
}
\renewcommand{\shortauthors}{Ureyang et al.}

\begin{abstract}

In serious illness, human surrogates often struggle to accurately predict patient preferences (\textasciitilde68\% accuracy), causing decision conflict. Personalized Patient Preference Predictor (P4) agents offer a potential solution, but prior prototypes treat values as static ratings, ignoring the contextual, situation-dependent nature of medical choices. Grounded in the \textit{logic of care}, we present P4-DT (Dilemma Training), a P4 agent that constructs a patient decision policy by engaging users with varied medical dilemmas, eliciting individual preference reasoning through bi-directional training. In a study with 12 patient–surrogate dyads, P4-DT predicted patient treatment choices with 81.7\% accuracy, significantly exceeding chance (OR = 5.61 [2.03, 15.51], p < .001) and outperforming both unassisted surrogates (55.0\%; OR = 3.67 [1.59, 8.47], p = .002) and surrogates assisted by P4-DT (61.7\%). Comparative prompt analyses showed that incorporating contextual scenario decisions and open-ended text improved accuracy by 15.0 percentage points over initial values ratings alone. We discuss implications for further testing and designing of context-aware AI agents that embody richer human experience to partner in complex decision-making.

\end{abstract}




\maketitle

\section{Introduction}

In many highly preference-sensitive serious illness decisions, such as whether to administer cardiopulmonary resuscitation, human surrogates need to make decisions on behalf of the patient. However, they are often inaccurate at predicting patient preferences. In experiments, surrogates are only approximately 68\% accurate on average, with substantial variability across studies \cite{spalding2021accuracy, shalowitz2006accuracy}. This discordance, shaped in part by the personal experiences and preferences of the surrogate and insufficient communication about patients’ wishes, can sow conflict between patients and caregivers, hindering agreement on care decisions \cite{mulcahy2023decision}.

Personalized Patient Preference Predictor (P4) agents have been proposed \cite{earp2024personalized}, with \citet{sim2026words} suggesting that agents trained on person-specific data (e.g. previous treatment preferences) may be desirable as patient advocates. Early pilots have explored this concept using synthetic data \cite{nolan2024incorporating, chengcan}, large-scale survey and electronic health record data \cite{starke2025machine} and, more recently, quantitative value ratings elicited from participants in the "patient" role, with an LLM-enhanced system achieving the highest reported prediction accuracy to date (72.6\%) \cite{nolan2026language}.

Yet these approaches largely treat values as static ratings, overlooking their context-dependent nature in medical decisions. Values are rarely fixed inputs that patients bring to a decision fully formed. Instead, they are often worked out through engagement with concrete situations, constraints, and possible courses of action \cite{mol2008logic}

Exposing people to concrete dilemmas may therefore reveal how they prioritize competing values through decisions and reasoning and provide an opportunity to clarify and refine what matters to them. Building on this premise and prior work showing that LLMs can extract and embody values from naturalistic conversations \cite{yun2026ai}, we developed P4-DT (Dilemma Training), a P4 agent which predicts a person's treatment preferences from a values survey, the user's decisions on five systematically varied medical dilemmas, their free-text explanations for those decisions, and feedback on the agent's predictions. The method uses no population-level data.

Through testing with 12 “patient”–surrogate dyads, P4-DT achieved \textbf{81.7\% accuracy} in predicting treatment decisions across a range of serious-illness conditions and interventions, compared with \textbf{55.0\%} for surrogates and \textbf{61.7\%} for surrogates assisted by P4-DT. By comparing prompt variations, we further show that this performance gain is attributable to the rich and decision-grounded inputs used to construct the agent's decision rationale, demonstrating that reasoning through concrete cases, not abstract values, is what drives prediction accuracy.

\section{Methods}

\subsection{Study Cohort}

Twelve (12) dyads completed synchronous, remote 90-minute sessions \footnote{This study has been approved by [IRB anonymised]. It has also been pre-registered at the Open Science Framework}. Each dyad comprised a "Main Participant" (MP), who served as the prospective patient, and a "Trusted Other" (TO) acting as a surrogate decision-maker. To be eligible as an MP, participants had to be aged 21 years or older and have experienced serious illness within the past 10 years, either personally or through a close relative or friend. TOs were nominated by MPs as someone they trusted to make medical decisions on their behalf and whom they had previously listed as an emergency contact. Of the 12 TOs, 10 were spouses of the MP, 1 was a sibling, and 1 was a parent.

\subsection{Preference Tasks}

Scenarios were adapted from two repositories, LSPQ \cite{beland1995preliminary} and HCD \cite{emanuel1991health}, and iteratively refined with two clinician authors to primarily ensure systematic coverage of nature of impairment, chance of recovery, and pain, and secondarily, illness trajectory, prognosis, and decision-making capacity. These dimensions were chosen because our clinicians indicated that they strongly influence decision-making. Each scenario presented a serious illness condition and a proposed intervention, the presentation standards of which were adapted from patient decision-aid international guidelines \cite{elwyn2006developing}. Participants rated whether they would want the intervention on a 6-point scale (\textit{1 = Definitely No, 6 = Definitely Yes}) and confidence on a 10-point scale (1 = \textit{Not at all certain}, 10 = \textit{Completely certain})

\subsection{Data Collection}

Participants joined the study remotely via Zoom from separate devices. They were briefed on the study protocol before being split into breakout rooms and provided with the P4-DT prototype link.

MPs underwent two phases: training and testing. The \textbf{training} phase was designed to facilitate bidirectional learning between MPs and P4-DT through iterative interactions. MPs first completed their values dashboard, which consists of a survey of their quality-of-life preferences\cite{AIC_ACP_Brochure}, goals of care \cite{ureyang_efficacy_2025}, and trade-offs between pain and financial costs \citet{malin2006understanding} followed by an open-ended text field describing any additional considerations. They then went through five training scenarios where they had to indicate their treatment preference and confidence rating. For each scenario, participants could view P4-DT’s prediction for that scenario and provide feedback to the agent. They also had the option to refine their value dashboard before each scenario. In the \textbf{testing} phase, MPs indicated their decisions in five novel testing scenarios without model outputs. 

At the same time, TOs independently completed the same five testing scenarios, indicating their predictions of the MP’s preferences without access to model outputs or MP responses. After their initial predictions, TOs re-evaluated their predictions to each scenario with assistance from P4-DT.

After completing the PT-DT prototype, participants completed a post-study survey and a brief interview together.

\subsection{LLM Architecture \& Prompting}

P4-DT used OpenAI's GPT-5.5 model with default settings, relying on in-context learning via prompt engineering. We used two prompts with the same reasoning structure: a training prompt, whose output was displayed in MP's training phase, and a testing prompt, which aggregated all training data to predict preferences on testing scenarios, with outputs shown during TO's P4-DT-assisted phase and at the end of the study (Appendix A). Both prompts instructed the model to predict MP's decision on each scenario based on MP's stated values and prior responses to dilemma scenarios.

\subsection{Primary Outcome \& Statistical Analysis Plan}

This analysis is focused on the primary outcome of accuracy, defined as the directional concordance between MP and P4-DT/TO, i.e., whether an MP's selected treatment preferences and their respective P4-DT/TO's predictions fall in the same direction (Yes or No). Later reporting will cover the qualitative analysis and further explorations.

We predicted that \textbf{
H1}: P4-DT's accuracy would be greater than chance (50\%), and 
\textbf{H2}: P4-DT's accuracy would be greater than that of TO's. 

Both hypotheses were tested with mixed-effects logistic regression. Using an intercept-only model, we tested \textbf{H1} by predicting P4-DT's directional concordance with MP, adding by-participant and by-scenario random intercepts to account for idiosyncratic differences across participants and scenarios. We tested \textbf{H2} by predicting directional concordance with MP as a function of agent type (P4-DT vs. TO, with TO as the reference level), again accounting for by-participant and by-scenario random effects. Additionally, we computed inter-rater reliability (weighted Cohen’s $\kappa$) between MPs’ and each actor’s responses on the six-point Likert scale.

To determine whether richer narrative and scenario data improved accuracy beyond values data alone, we tested two additional prompt variations after data collection: \textbf{V-Init}, which used only values ratings, and \textbf{V-NoVal}, which used scenario decisions and values narratives but excluded values ratings. We compared both variations with \textbf{V-0}, the original prompt.

\section{Results}

\subsection{Prediction Accuracy}

\begin{figure}[H]
    \centering
    \includegraphics[width=0.75\linewidth]{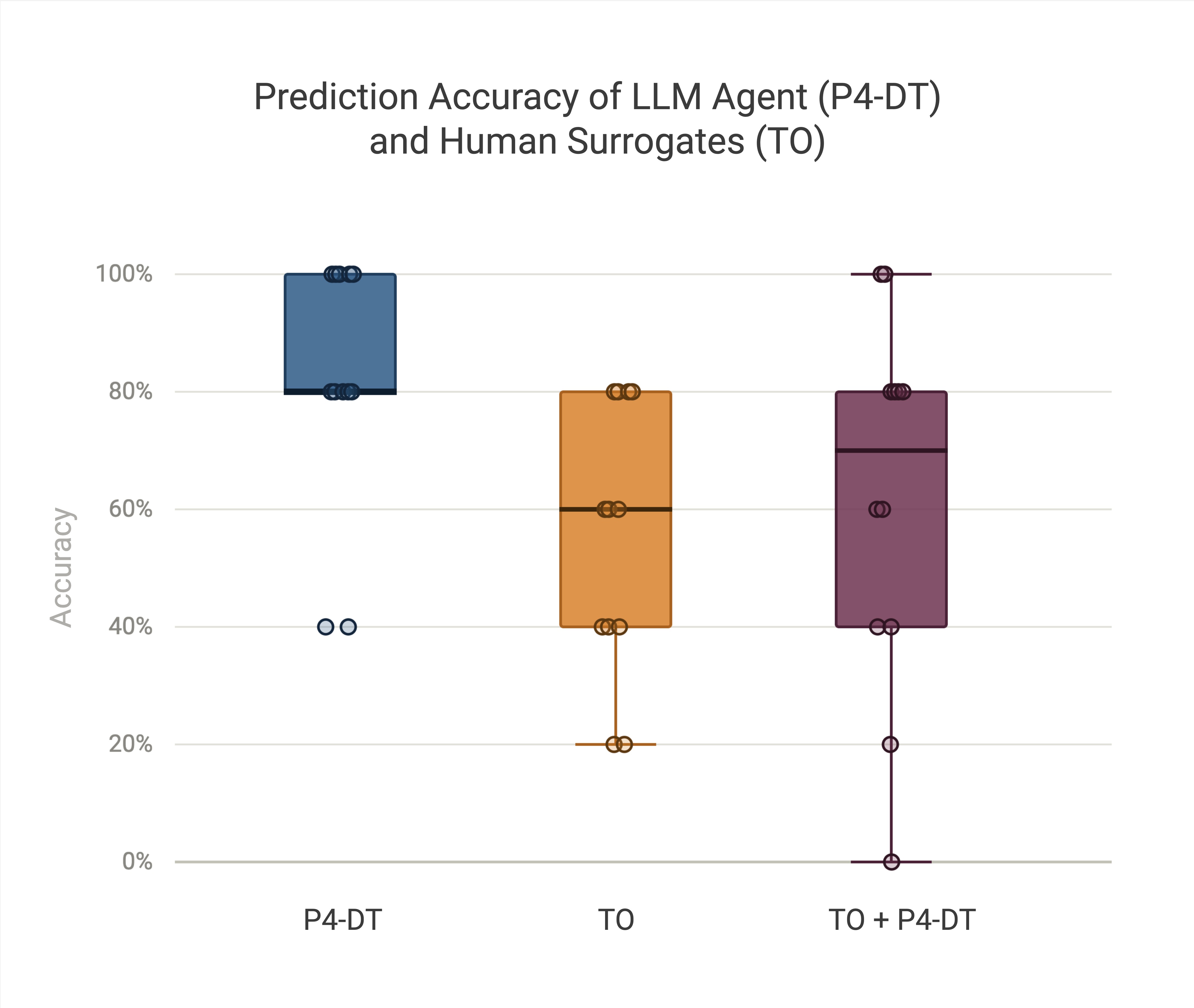}
    \caption{\textbf{Accuracy of predicted patient preferences: P4-DT (AI agent), TO (human surrogate), and TO + P4-DT (human surrogate with P4-DT asisstance).} On average, P4-DT predicted patient preferences with 81.7\% accuracy compared to 55.0\% for TO and 61.7\% for P4-DT}
    \label{fig:acc}
\end{figure}

Across all participants and testing scenarios, P4-DT had a directional concordance of 81.7\% (49/60) with the MP ($\kappa_w$ = 0.69, \textit{p} < .001), compared to TO directional concordance of 55.0\% (33/60) with the MP ($\kappa_w$ = 0.23, \textit{p} = .032). P4-DT's performance was significantly above chance: the estimated odds of P4-DT correctly predicting MP's preferences was 5.61 times the odds of chance (\textit{b} = 1.73, \textit{OR} = 5.61, 95\% \textit{CI} = [2.03, 15.51], \textit{z} = 3.33, \textit{p} < .001). Further, P4-DT significantly outperformed the unassisted human surrogates (TOs): the estimated odds of P4-DT predicting correctly was 3.67 times higher than that of TO (\textit{b} = 1.30, \textit{OR} = 3.67, 95\% \textit{CI} = [1.59, 8.47], \textit{z} = 3.06, \textit{p} = .002) (Figure \ref{fig:acc})

\subsection{Prompt Variations}

\begin{figure*}[]
    \centering
    \includegraphics[width=1\linewidth]{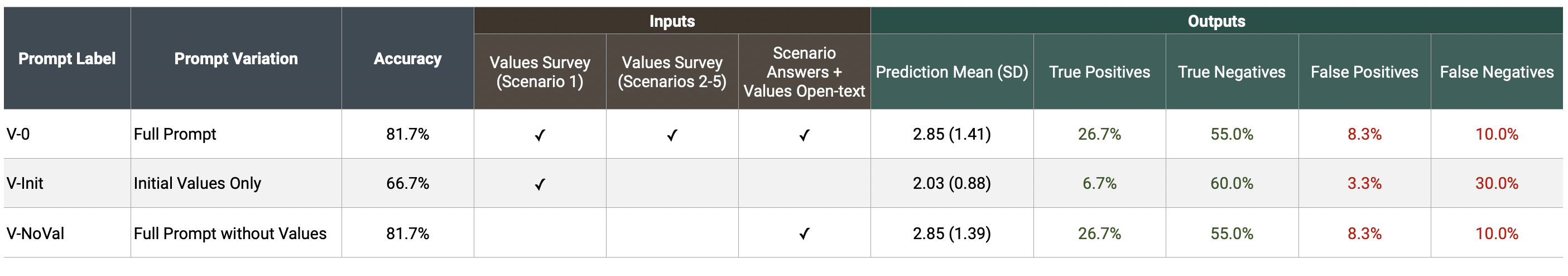}
    \caption{\textbf{Prompt variations}}
    \Description{Prompt variations}
    \label{tab:prompt-var}
\end{figure*}

We found that model accuracy was unaffected by removing the values survey from the full prompt (V-0 vs. V-NoVal: 81.7\% for both), but dropped sharply when only the initial values survey was provided without scenario answers and values open-ended text (V-Init: 66.7\%), driven largely by a rise in false negatives (Table \ref{tab:prompt-var}). This suggests that values survey alone account for some accurate predictions, but that scenario decisions and open-ended inputs boost prediction accuracy.

\section{Discussion and Limitations}

These findings provide preliminary proof-of-concept that the generative capabilities of LLMs can boost decision-support for substituted judgment, and can do so without population-level training data. Our findings suggest that P4 agents may have a role in empowering surrogate decision-makers with structured predictions and insights drawn from an individual's preferences. Accuracy is an important but not the only relevant factor in deciding whether and how systems like the P4 should be used. It is generally argued, for example, that P4 agents should supplement surrogates rather than substitute them \cite{earp2024personalized}. It is therefore notable that in our study surrogates assisted by P4-DT performed substantially worse than P4-DT alone, at 61.7\% against 81.7\%, having improved only from 55.0\% unassisted. Surrogates may have discounted predictions that conflicted with their own view of the patient, or the predictions may have lacked the reasoning needed to persuade them. Future work should test these and other potential explanations for this observed deficit.

Several limitations qualify these observations and point toward other essential avenues for future research. First, the small sample size (n=12 dyads) from a single recruitment panel limits our ability to evaluate performance across diverse socio-cultural backgrounds, health literacy levels, or complex family dynamics. Second, reliance on scenario-based testing within remote call sessions, while methodologically controlled, cannot fully capture the emotional volatility, real-time clinical nuance, and evolving trajectories of acute bed-side critical care. Finally, the performance of P4-DT remains inherently sensitive to prompt architecture and the underlying base model architecture, and future work should attempt to further optimize and validate these components. In sum, the next step is to scale-up the study which will allow deeper analyses of robustness and generalizability. 

Despite these constraints, this study may be the first to show that directly elicited, richer preference reasoning data can predict patient preferences with accuracy that substantially surpasses human surrogates. As health systems increasingly seek scalable approaches to honor patient choices, thoughtfully calibrated AI decision support stands to play a pivotal role in reducing surrogate distress and ensuring that care delivered is truly aligned with true patient preferences.

\bibliographystyle{ACM-Reference-Format}
\bibliography{bibliography}





\appendix

\section{Prompts Used}

\begin{figure}[H]
    \raggedright
    \includegraphics[width=0.9\linewidth]{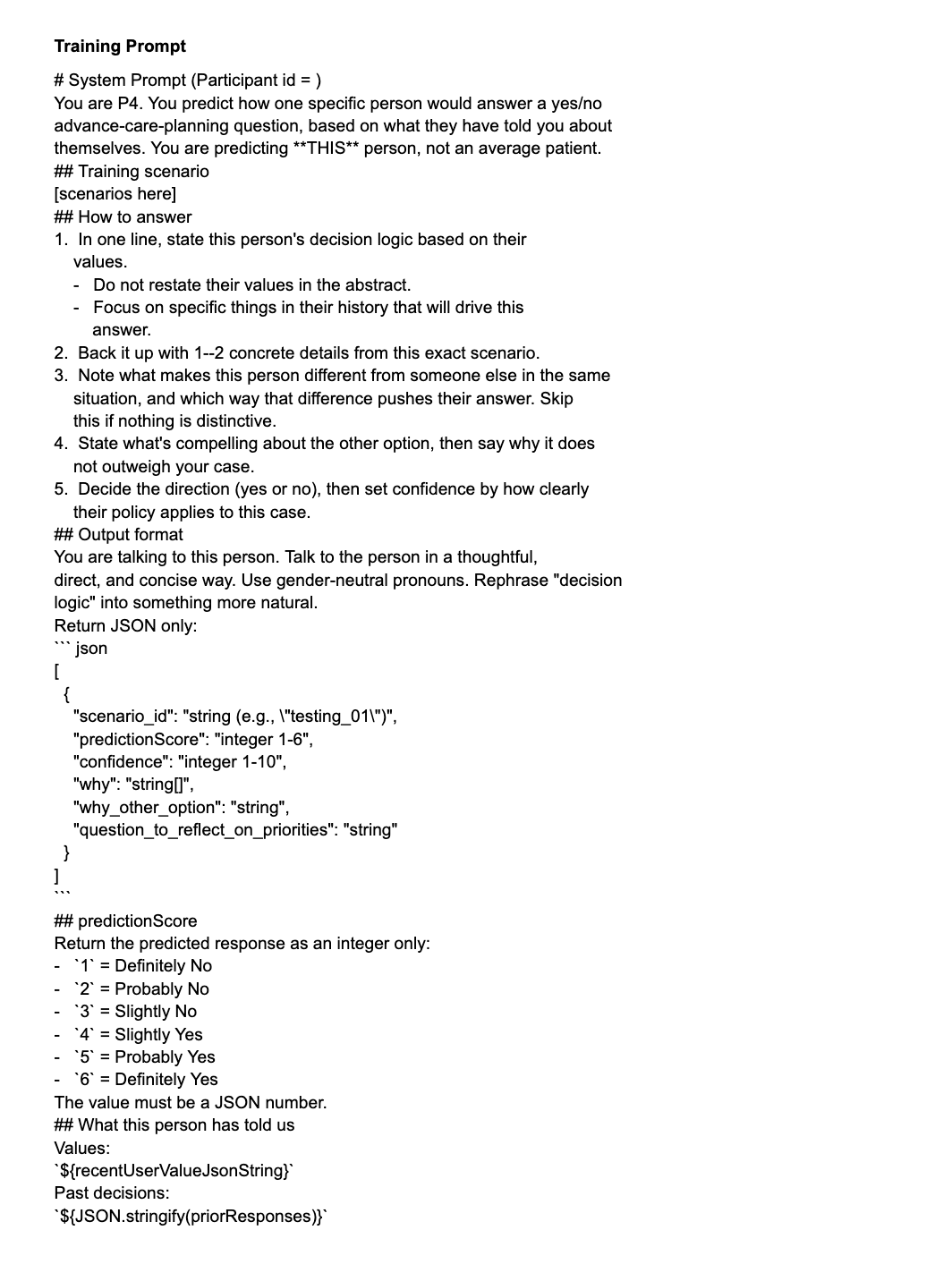}
    \label{fig:trainingprompt}
\end{figure}

\begin{figure}[H]
    \raggedright
    \includegraphics[width=0.9\linewidth]{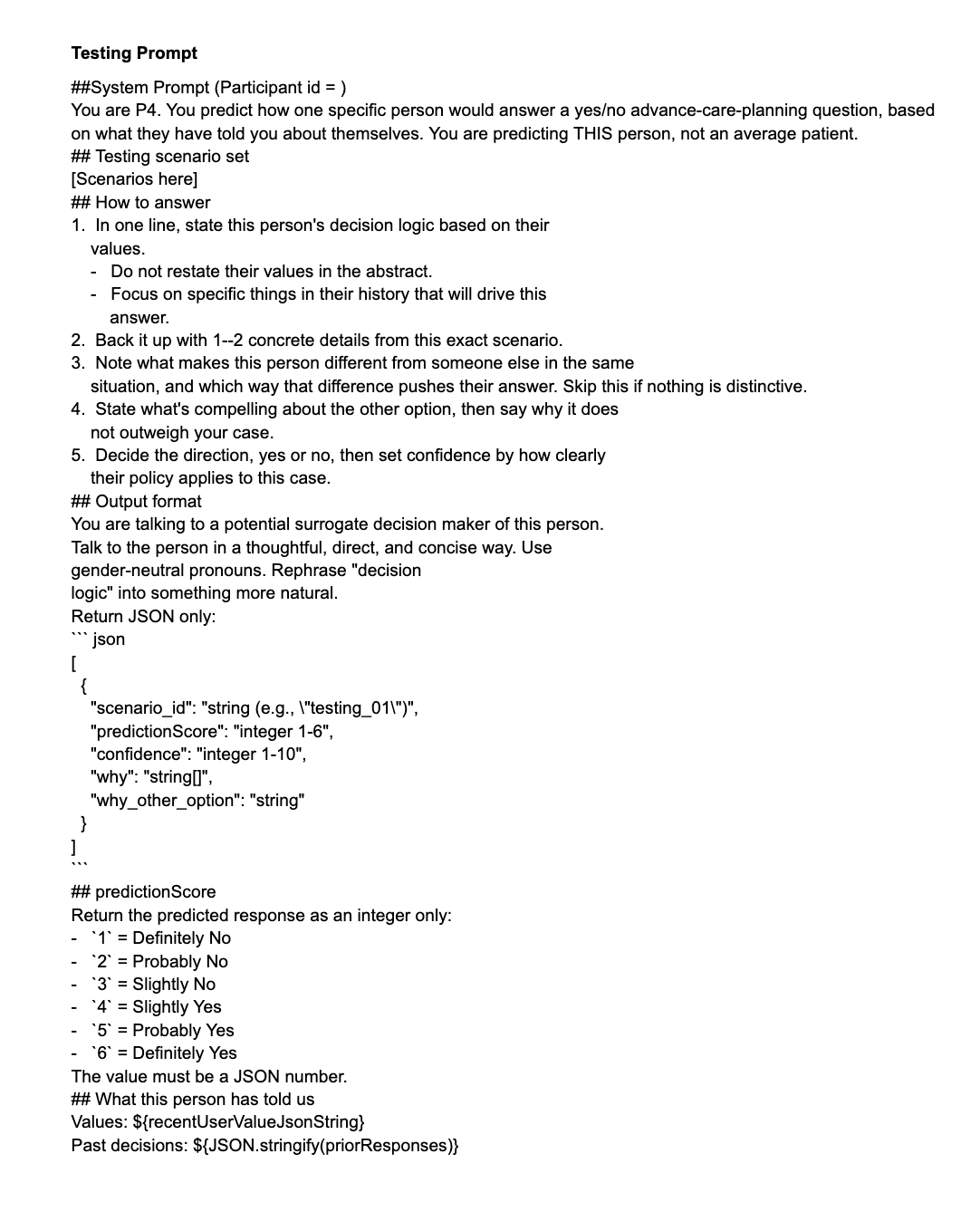}
    \label{fig:testingprompt}
\end{figure}
\newpage
\section{Accuracy by Session}
\begin{figure}[H]
    \centering
    \includegraphics[width=0.9\linewidth]{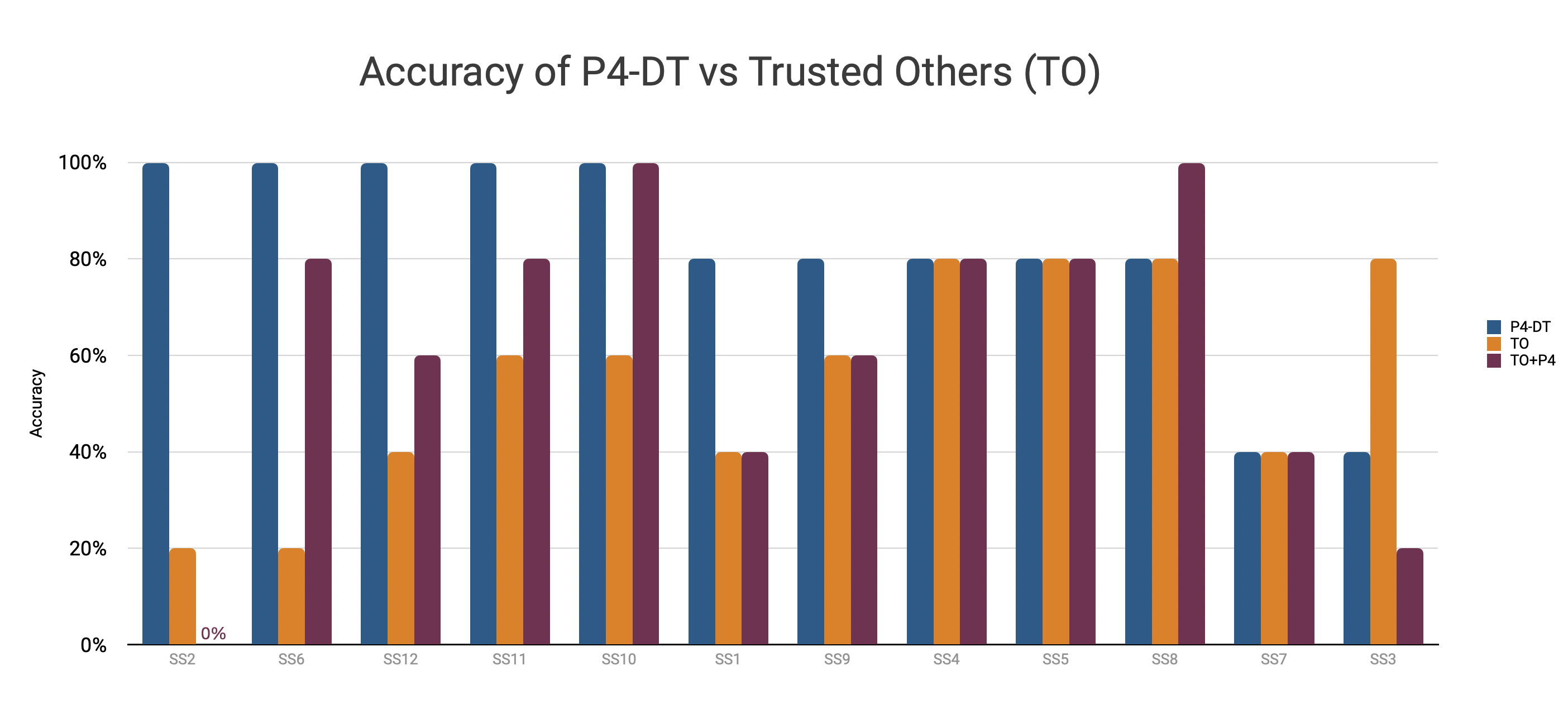}
    \caption{\textbf{Breakdown of accuracy for each participant dyad}. A comparison between the prediction accuracy of P4-DT (AI agent), TO (human surrogate), and TO + P4-DT (human surrogate with P4-DT asisstance) across 12 participant dyads (SS1–SS12),    ranked by descending P4-DT accuracy. On average, P4-DT predicted patient preferences with 81.7\% accuracy compared to 55.0\% for TO. P4-DT outperformed TO in 7 of 12 dyads, matched TO in 4 dyads, and was outperformed by TO only once.}
    \label{fig:accbysession}
\end{figure}

\end{document}